\RequirePackage[l2tabu, orthodox]{nag}
\documentclass{scrartcl}
\usepackage{typearea}
\typearea{12}
\usepackage{amssymb, amsmath}
\usepackage{enumitem}
\usepackage[all, warning]{onlyamsmath}
\usepackage[dvips]{graphicx}
\usepackage{tikz}
\usetikzlibrary{arrows}
\usepackage{color}
\usepackage{authblk}
\usepackage{url}
\newtheorem{THEO}{Theorem}[section]
\newtheorem{ALGo}[THEO]{Algorithm}
\newtheorem{CONJ}[THEO]{Conjecture}
\newtheorem{COND}[THEO]{Condition}
\newtheorem{CORO}[THEO]{Corollary}
\newtheorem{DEFI}[THEO]{Definition}
\newtheorem{EXAMP}[THEO]{Example}
\newtheorem{FACT}[THEO]{Fact}
\newtheorem{HYPO}[THEO]{Hypothesis}
\newtheorem{LEMM}[THEO]{Lemma}
\newtheorem{PROB}[THEO]{Problem}
\newtheorem{PROP}[THEO]{Proposition}
\newtheorem{REMA}[THEO]{Remark}
\newcommand{\theo}{\begin{THEO}}
\newcommand{\algo}{\begin{ALGo} \rmfamily}
\newcommand{\cond}{\begin{COND}}
\newcommand{\conj}{\begin{CONJ}}
\newcommand{\coro}{\begin{CORO}}
\newcommand{\defi}{\begin{DEFI} \rmfamily}
\newcommand{\examp}{\begin{EXAMP} \rmfamily}
\newcommand{\fact}{\begin{FACT}}
\newcommand{\hypo}{\begin{HYPO} \rmfamily}
\newcommand{\lemm}{\begin{LEMM}}
\newcommand{\prob}{\begin{PROB} \rmfamily}
\newcommand{\prop}{\begin{PROP}}
\newcommand{\rema}{\begin{REMA} \rmfamily}
\newcommand{\etheo}{\end{THEO}}
\newcommand{\ealgo}{\end{ALGo}}
\newcommand{\econd}{\end{COND}}
\newcommand{\econj}{\end{CONJ}}
\newcommand{\ecoro}{\end{CORO}}
\newcommand{\edefi}{\end{DEFI}}
\newcommand{\eexamp}{\end{EXAMP}}
\newcommand{\efact}{\end{FACT}}
\newcommand{\ehypo}{\end{HYPO}}
\newcommand{\elemm}{\end{LEMM}}
\newcommand{\eprob}{\end{PROB}}
\newcommand{\eprop}{\end{PROP}}
\newcommand{\erema}{\end{REMA}}

\DeclareMathOperator{\conv}{conv}
\DeclareMathOperator{\ext}{ext}

\makeatletter
\global\let\tikz@ensure@dollar@catcode=\relax
\makeatother

\title{Boundary Modeling in Model-Based Calibration for Automotive Engines via the Vertex Representation of the Convex Hulls}
\author[1]{Hayato Waki\thanks{744 Motooka, Nishi-ku, Fukuoka 819-0395, Japan. waki@imi.kyushu-u.ac.jp}}
\author[2]{Florin Nae\thanks{16/F KDX Sakuradori Bldg., 3-20-17, Marunouchi, Naka-ku, Nagoya, Aichi, 460-0002 Japan. Florin.Nae@mathworks.co.jp}}
\affil[1]{Institute of Mathematics for Industry, Kyushu University}
\affil[2]{The MathWorks G.K. Nagoya Office}
\date{First version : April 26, 2016, Revised : May 11, 2016}

\begin{document}
\maketitle

\begin{abstract}
When using the convex hull approach in the boundary modeling process, Model-Based Calibration (MBC) software suites -- such as Model-Based Calibration Toolbox from MathWorks -- can be computationally intensive depending on the amount of data modeled. The reason for this is {that} the half-space representation of the convex hull {is used}. We discuss here  another representation of the convex hull, the vertex representation, which proves capable to  reduce the computational cost. Numerical comparisons in this article are executed in MATLAB by using MBC Toolbox commands, and  show that for certain conditions, the vertex representation outperforms the half-space representation.    
\end{abstract}

\section{Introduction}\label{intro}

Model-Based Calibration ({\itshape abbr.} MBC) is a systematic approach for more cost-effective and short-term development of automotive engines, that enables engineers to design more efficient automotive engines, {\itshape e.g.}, more {fuel-efficient} and/or eco-friendly engines. For efficient design of automotive engines, mathematical models for automotive engines are created in MBC, and statistics and optimization are applied  {to} the model by using MBC software, such as \cite{mbctool}. 

Boundary modeling is one of the processes in MBC used to represent/approximate a region {where} the automotive engine works normally, {\itshape e.g.}, without misfire and knock of the engine. We call the region {\itshape the admissible operation domain} ({\itshape abbr.} AOD). 
In general, {as} it is assumed that  internal-combustion engines are highly nonlinear systems, it is impossible to exactly represent the AOD of the automotive engine from {a} finite number of acquired data. {Thus} one approximates the AOD instead of representing it exactly.  One of the approximations of AOD is to use the convex hull of a set of data. This is a simple way to approximate AOD from data and is implemented in MBC software, such as \cite{mbctool}. In addition to the convex hull, the use of support vector machine for the approximation of AOD is also proposed in \cite{kieft2014}. 

An AOD is used as a constraint in constrained optimization problems. One can assume that  some of optimal solutions will lie on the boundary of the feasible region, otherwise the constraints would be irrelevant. That is why a proper handling of AODs is important in engine optimization problems. 

The motivation of this article comes from the comment in  \cite{workshop} that some of the MBC software suites spend much computational time constructing a convex hull boundary model. In general, two representations for the convex hull of a set of points are possible, {\itshape the half-space representation} and {\itshape the vertex representation}. The reason for the comment was {that} the half-space representation  for the convex hull of a set of points {is typically used} by software like MBC Toolbox, instead of the vertex representation. 

The contribution of this article is to propose the use of the convex hull in the vertex representation instead of the half-space representation.  
In practice, the former representation seems to perform better than the latter. 
In fact, the numerical comparison in this article shows that the vertex representation is less computationally intensive than the hyperplane representation in the case {when} the dimension of inputs for engine models is more than five. 

The organization of this article is as follows: convex hull modeling theory is discussed in Section \ref{preliminary}. Section \ref{application} provides an application of the vertex representation of the convex hull and numerical experiments. {Conclusion is} given in Section \ref{conclusion}. Throughout this article, we assume that the measured engine data was  acquired by keeping the engine under test at  steady  condition by controlling its inputs. 

\section{Preliminaries}\label{preliminary}

We give a brief introduction on boundary modeling via the convex hull in Section \ref{bm}, and some definitions and facts on  the convex hull for a set of points in Section \ref{convexhull}. Refer to \cite{barvinok, branko} for more details regarding the convex hull mathematical representation. 
\subsection{Boundary modeling in model-based calibration}\label{bm}

The behavior of automotive engines is represented by  the state space representation. One of the simplest formulations is  as follows: 
\[
\left\{
\begin{array}{rcl}
\displaystyle\frac{dx}{dt} &=& f(x, u), \\
y &=& g(x, u), 
\end{array}
\right. 
\]
where $t$ is time, $x$, $u$ and $y$ are vectors which represent the state of the automotive engine, the input signals into the engine and the output signals from the engine, respectively. 

Control theory, statistics and optimization are applied to such mathematical models of automotive engines to design more {fuel-efficient} and/or eco-friendly engines. MBC is a systematic approach for aiding such an efficient design of automotive engines and consists of some processes, such as the design of experiments and the response surface methodology. 

Boundary modeling is a functionality used in MBC, and is applied to define an AOD for a mathematical engine model. Input signals for automotive engines under development have specific operating ranges and dynamics. In addition, automotive engines may not behave normally when some specific {input} signals are used, leading to undesirable events such as misfire and knock of the engine. In boundary modeling,  one defines that  approximates/represents a region of input signals where automotive engines  behave normally, {\itshape e.g.}, without misfire and knock of the engine.  

\begin{figure}[htbp]
\begin{minipage}{0.5\hsize}
\centering
\begin{tikzpicture}
\draw[->,>=latex'] (0,0.5) -- node[pos=0.5,below,rotate=0]{Input 1} (4.5,0.5);
\draw[->,>=latex'] (0,0.5) -- node[pos=0.5,above,rotate=90]{Input 2} (0,4.5);
\filldraw[blue!25,draw=black]  (1,2) -- (0.5,3) -- (2,4) -- (3.5, 4) -- (3,1) -- cycle;
  \foreach \point in {(1,2), (0.5,3), (2,4), (3.5, 4), (3,1), (1.4, 3.5), (3, 3)} {
    \fill[black] \point circle[radius=2pt];
  }
    \foreach \point in {(1,1.75), (0.3,3), (1.8,4.25), (3.75, 4.1), (3.1,0.8), (4, 2.5), (1, 3.75)} {
    \fill[red] \point circle[radius=2pt];
  }
\end{tikzpicture}
\end{minipage}
\begin{minipage}{0.5\hsize}
\centering
\begin{tikzpicture}
\draw[->,>=latex'] (0,0.5) -- node[pos=0.5,below,rotate=0]{Input 1} (4.5,0.5);
\draw[->,>=latex'] (0,0.5) -- node[pos=0.5,above,rotate=90]{Input 2} (0,4.5);
\filldraw[blue!25,draw=black]  (1,2) -- (0.5,3) -- (2,4) -- (3.5, 4) -- (3,1) -- cycle;
  \foreach \point in {(1,2), (0.5,3), (2,4), (3.5, 4), (3,1), (1.4, 3.0), (2.7, 2.75)} {
    \fill[black] \point circle[radius=2pt];
  }
    \foreach \point in {(1,1.75), (0.3,3), (1.8,4.25), (3.75, 4.1), (3.1,0.8), (2.8, 2.5), (1.3, 3.2)} {
    \fill[red] \point circle[radius=2pt];
  }
\end{tikzpicture}
\end{minipage}
\caption{{Examples} of the approximation of the AOD by the convex hull}
\label{fig:aod}
\end{figure}
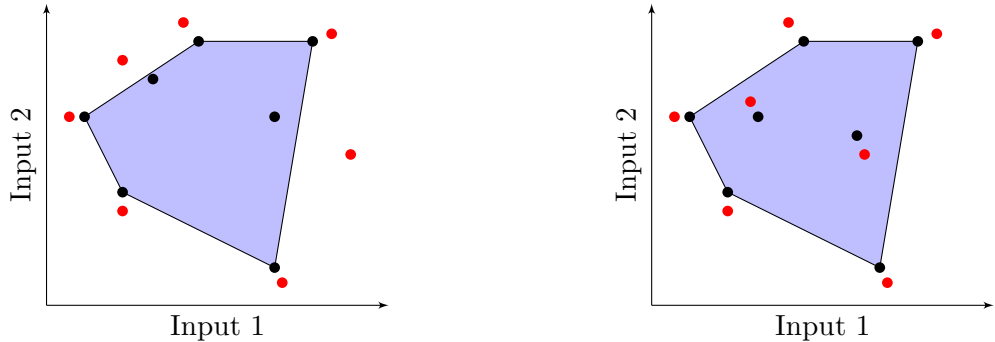

One of the approximations of the AOD is the convex hull of a set of a finite number of input signals {by which} the automotive engine behaves normally. This approximation may be too rough, but  is a simple way to define an AOD in practice. In fact, it is implemented in some MBC software, such as \cite{mbctool}. Figure \ref{fig:aod} displays examples of the approximation of the AOD by the convex hull. In Figure \ref{fig:aod}, black circles are input signals {by which} the automotive engine behaves normally, and red circles indicates input signals {by which} the automotive engine does not behave normally. The blue region is the approximation of the AOD via the convex hull.

Note that as we mentioned, the approximation of the AOD by the convex hull may be rough. In fact, it does not always represent the region where the automotive engine behave normally. For instance,  the approximation at the right of Figure \ref{fig:aod} contains red circles, which means that  the automotive engine does not behave normally around the circle. 


The approximation of the AOD is used  in other processes in MBC  as follows:    
\begin{enumerate}[label=(P\arabic*)]
\item Problem of determining  whether  a new point is in the approximated AOD or not. This is mathematically formulated as the problem of determining  
\[
\hat{v} \in P \mbox{ or } \hat{v} \not\in P, 
\]
where $\hat{v}$ is a new point and $P$ is an approximation of the AOD. \label{P1}
\item Optimization of some objective functions over the  approximated AOD or a subset of the AOD for more realistic situation in response surface methodology. This is mathematically formulated as 
\[
\displaystyle\min_{v\in\mathbb{R}^n}\left\{f(v) : g_j(v) \ge 0 \ (j=1, \ldots, k), v\in  P\right\}, 
\]
where $f(v)$ is the objective function and $g_j(v) \ge 0$ is an engine operating  constraint. \label{P2}
\end{enumerate}

\subsection{Convex hull for a set of points in $\mathbb{R}^n$}\label{convexhull}

Let $V=\{v_1, \ldots, v_m\}$ be a finite set of distinct points in $\mathbb{R}^n$. A point
\[
x = \sum_{i=1}^m \alpha_i v_i, \mbox{where } \sum_{i=1}^m\alpha_i=1, \alpha_i\ge 0\mbox{ for } i=1, \ldots, m, 
\] 
is called a {\itshape convex combination} of $v_1, \ldots, v_m$. In particular, the set $\{\alpha a + (1-\alpha) b :  0\le \alpha\le 1\}$ is called the {\itshape line segment with the endpoints $a$ and $b$} and denoted by $[a, b]$. 

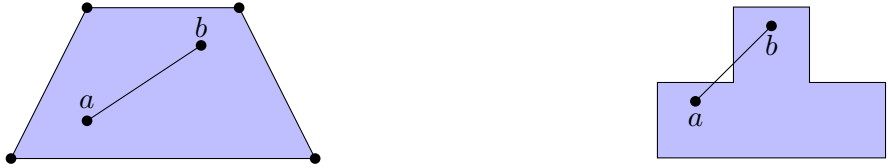
\begin{figure}[htbp]
\begin{tabular}{c}
\begin{minipage}{0.5\hsize}
\centering
\begin{tikzpicture}
\filldraw[blue!25,draw=black] (-1,0) -- (0,2) -- (2,2) -- (3,0) -- cycle;
  \foreach \point in {(-1,0), (0,2), (2,2), (3,0)} {
    \fill[black] \point circle[radius=2pt];
  }
  \foreach \point in {(0, 0.5), (1.5,1.5)} {
    \fill[black] \point circle[radius=2pt];
  }
  \draw (0, 0.5) -- (1.5,1.5);
  \node[circle] (a) at (0, 0.75) {$a$};
   \node[circle] (b) at (1.5,1.75) {$b$};
\end{tikzpicture}
\end{minipage}
\begin{minipage}{0.5\hsize}
\centering
\begin{tikzpicture}
\filldraw[blue!25,draw=black] (0,0) -- (0,1) -- (1,1) -- (1,2) -- (2,2) -- (2,1) -- (3, 1) -- (3,0) -- cycle;
  \foreach \point in {(0.5, 0.75), (1.5,1.75)} {
    \fill[black] \point circle[radius=2pt];
  }
  \draw (0.5, 0.75) -- (1.5,1.75);
  \node[circle] (a) at (0.5, 0.5) {$a$};
   \node[circle] (b) at (1.5,1.5) {$b$};
\end{tikzpicture}
\end{minipage}
\end{tabular}
\caption{Convex set (left) and nonconvex set (right)}
\label{fig:example1}
\end{figure}

A set $K\subseteq\mathbb{R}^n$ is {\itshape convex} if for every $a, b\in K$, the line segment $[a, b]$ is contained in $K$. We define the empty set $\emptyset$ as a convex set. Figure \ref{fig:example1} displays an example of convex and nonconvex sets. In fact, for the set at the left of Figure \ref{fig:example1}, we see that for every $a, b$ in the set, the line segment $[a, b]$ is contained in the set, which implies that the set is convex. In contrast, the line segment $[a, b]$ is not contained in the set at the right of Figure \ref{fig:example1}.

Let $K\subseteq\mathbb{R}^n$ be a convex set. A point $x\in K$ is an {\itshape extreme point} or {\itshape vertex} of $K$ if $y, z\in K$, $0<\alpha<1$ and $x=\alpha y + (1-\alpha)z$ imply $x=y=z$. In other words, the extreme point of $K$ is a point which does not have any convex combinations with other points in $K$. For instance, at the set of the left in Figure \ref{fig:example1}, the black circles at the corners indicate an extreme point of the convex set. We denote the set of extreme points in $K$ by $\ext(K)$.  

\begin{figure}[htbp]
\centering
\begin{tikzpicture}
\filldraw[blue!25,draw=black] (0, 0) -- (2, 0) -- (3, 2)  -- (0, 1) -- cycle;
  \foreach \point in {(0, 0), (2, 0), (3, 2), (1, 1), (0, 1)} {
    \fill[black] \point circle[radius=2pt];
  }
    \node (a) at (0.0, -0.4) {$(0, 0)$};
   \node (b) at (2.0, -0.4) {$(2, 0)$};
   \node (c) at (3.6, 2.0) {$(3, 2)$};
   \node (d) at (0.0, 1.4) {$(0, 1)$};
   \node (e) at (1.0, 0.6) {$(1, 1)$};
\end{tikzpicture}
\caption{Convex hull of $V=\{(0, 0), (2, 0), (3, 2), (1, 1), (0, 1)\}$}
\label{fig:example2}
\end{figure}
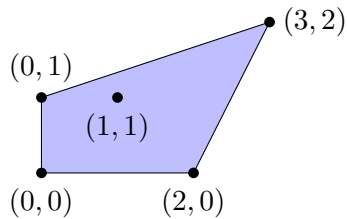

The convex hull $\conv(A)$ of a subset $A\subseteq\mathbb{R}^n$ is the set of all convex combination of points from $A$.  
%
For a set $V=\{v_1, \ldots, v_m\}$ of distinct points in $\mathbb{R}^n$, $\conv(V)$ is formulated mathematically as
\[
\conv(V) = \left\{
v\in\mathbb{R}^n : v = \sum_{i=1}^m\alpha_i v_i \mbox{ for some }\sum_{i=1}^m\alpha_i = 1, \alpha_i\ge 0 \ (i=1, \ldots, m) 
\right\}. 
\]
Since some points in $V$ are extreme points of the convex hull, this representation of $\conv(V)$  is called the {\itshape vertex representation} ({\itshape abbr.} V-representation).  Figure \ref{fig:example2} displays an example of the convex hull of $V = \{(0, 0), (2, 0), (3, 2), (1, 1), (0, 1)\}$. Since all points except for $(1, 1)$ are extreme points, $\ext(\conv(V)) = \{(0, 0), (2, 0), (3, 2), (0, 1)\}$.  In fact, $(1, 1)$ is not the extreme point of the convex hull because $(1, 1)$ can be represented by a convex combination with $(2, 0)$, $(3, 2)$ and $(0, 1)$.  In addition, we see 
$\conv(V) = \conv(\ext(V))$ in Figure \ref{fig:example2}. 

A bounded convex set $K\subseteq\mathbb{R}^n$ is a {\itshape polytope} if $\ext(K)$ is a finite set. Clearly the convex hull of a set of a finite numbers of points in $\mathbb{R}^n$ is a polytope. A half-space is a set  which is defined as $\{x\in\mathbb{R}^n : a^Tx \le b\}$, with suitable $a\in\mathbb{R}^n$ and $b\in\mathbb{R}$. A set $P$ is called {\itshape polyhedron} if $P$ is formed as the intersection of finitely many half-spaces, {\itshape i.e.}, there exist $a_1, \ldots, a_k\in\mathbb{R}^n$ and $b_1, \ldots, b_k\in\mathbb{R}$ such that $P =\left\{
x \in \mathbb{R}^n : a_i^Tx \le b_i \ (i=1, \ldots, k)
\right\}$. 

Minkowski-Weyl's theorem ensures that every polytope can be reformulated as a polyhedron. This implies that one can describe the convex hull of a set of points by some half-spaces in addition to the V-representation, which  is called the {\itshape half-space representation} ({\itshape abbr.} H-representation).  

\theo\label{mw} (Minkowski-Weyl) 
Every polytope is polyhedron, {\itshape i.e.}, for a given polytope $P$,  there exist $a_1, \ldots, a_k\in\mathbb{R}^n$ and $b_1, \ldots, b_k\in\mathbb{R}$ such that $P=\{x\in\mathbb{R}^n : a_i^Tx\le b_i \ (i=1, \ldots, k)\}$. Moreover, every bounded polyhedron is also polytope, {\itshape i.e.}, for a given polyhedron $P$, there exist $v_1, \ldots, v_m\in P$ such that $P=\conv(V)$, where $V=\{v_1, \ldots, v_m\}$. 
\etheo 

 We give two examples of the V- and H-representations. We see from these examples that one needs to choose a suitable representation of the convex hull from the viewpoint of computation. 
  
 \examp\label{cube} ($n$-dimensional unit cube) Let $P = \{x\in\mathbb{R}^n : 0\le x_i\le 1 \ (i=1, \ldots, n)\}$. $P$ is called the $n$-{\itshape dimensional unit cube}. Figure \ref{fig:example3} displays an example of $3$-dimensional unit cube. This  is already the H-representation. In fact, we define $a_i\in\mathbb{R}^n, b_i\in\mathbb{R}$ \ $(i=1, \ldots, 2n)$ as follows:
 \[
 a_i = \left\{
 \begin{array}{cl}
 e_i & (i=1, \ldots, n), \\
 -e_i & (i=n+1, \ldots, 2n),
 \end{array}
 \right. \mbox{ and }  b_i = \left\{
 \begin{array}{cl}
 1 & (i=1, \ldots, n), \\
 0 & (i=n+1, \ldots, 2n),
 \end{array}
 \right.
 \]
 where $e_i$ is the $i$th $n$-dimensional standard unit vector. Then $P$ can be reformulated by $\{x \in \mathbb{R}^n : a_i^Tx\le b_i \ (i=1, \ldots, 2n)\}$. On the other hand, {for} $\ext(P) = \{x\in\mathbb{R}^n : x_i = 0 \mbox{ or } 1\}$,  the V-representation of $P$ is
 \[
 P = \conv(\{(0, 0, \ldots, 0), (1, 0, \ldots, 0), (0, 1, \ldots, 0), \ldots, (1, 1, \ldots, 1)\}). 
 \]
 We remark that the V-representation of $P$ needs $2^n$ extreme points in $\ext(P)$, whereas the H-representation needs only $2n$ half-spaces.  
 \eexamp

 \examp\label{cross} (Cross-polytope) 
 Let $P = \{x\in\mathbb{R}^n : |x_1|+\cdots +|x_n|\le 1\}$. $P$ is called the $n$-{\itshape dimensional cross-polytope}. Figure \ref{fig:example3} displays an example of the $3$-dimensional cross-polytope. The H-representation of $P$ is 
 \[
 P = \left\{x\in\mathbb{R}^n : \begin{array}{lcl}
 x_1+x_2+\cdots+x_n&\le& 1\\
 -x_1+x_2+\cdots+x_n&\le& 1\\
 x_1-x_2+\cdots+x_n&\le& 1\\
 -x_1-x_2+\cdots+x_n&\le& 1\\
 &\vdots&\\
  -x_1-x_2-\cdots-x_n&\le& 1\\
 \end{array}
\right\}. 
 \]
 Here the H-representation is the intersection of $2^n$ half-spaces. In contrast, the V-representation of $P$ can be formulated by $2n$ points in $\mathbb{R}^n$. In fact, since both $e_i$ and $-e_i$ are extreme points in $P$, the V-representation of $P$ is $P = \conv(\{\pm e_1, \ldots, \pm e_n\})$. 
 \eexamp
 \newcommand{\Depth}{2}
\newcommand{\Height}{2}
\newcommand{\Width}{2}

\begin{figure}[htbp]
\centering
\begin{tabular}{c}
\begin{minipage}{0.5\hsize}
\begin{tikzpicture}
\coordinate (O) at (0,0,0);
\coordinate (A) at (0,\Width,0);
\coordinate (B) at (0,\Width,\Height);
\coordinate (C) at (0,0,\Height);
\coordinate (D) at (\Depth,0,0);
\coordinate (E) at (\Depth,\Width,0);
\coordinate (F) at (\Depth,\Width,\Height);
\coordinate (G) at (\Depth,0,\Height);

\draw[black,fill=blue!25] (O) -- (C) -- (G) -- (D) -- cycle;
\draw[black,fill=blue!25] (O) -- (A) -- (E) -- (D) -- cycle;
\draw[black,fill=blue!25] (O) -- (A) -- (B) -- (C) -- cycle;
\draw[black,fill=blue!25,opacity=0.7] (D) -- (E) -- (F) -- (G) -- cycle;
\draw[black,fill=blue!25,opacity=0.7] (C) -- (B) -- (F) -- (G) -- cycle;
\draw[black,fill=blue!25,opacity=0.7] (A) -- (B) -- (F) -- (E) -- cycle;
\node (A) at (-0.75,\Width+0.15,0) {$(0, 1, 1)$};
\node (B) at (-0.75,\Width,\Height) {$(0, 0, 1)$};
\node (C) at (-0.75,0,\Height) {$(0, 0, 0)$};
\node (D) at (\Depth+0.75,0,0) {$(1, 1, 0)$};
\node (E) at (\Depth+0.75,\Width,0) {$(1, 1, 1)$};
\node (G) at (\Depth,0-0.5,\Height) {$(1, 0, 0)$};
\end{tikzpicture}
\end{minipage}
\begin{minipage}{0.5\hsize}
 \begin{tikzpicture}
\coordinate (A1) at (0,0,-1.75);
\coordinate (A2) at (-1.75,0,0);
\coordinate (A3) at (0,0,1.75);
\coordinate (A4) at (1.75,0,0);
\coordinate (B1) at (0,1.75,0);
\coordinate (C1) at (0,-1.75,0);

\draw [black, fill=blue!25] (A1) -- (A2) -- (B1) -- cycle;
\draw [black, fill=blue!25] (A4) -- (A1) -- (B1) -- cycle;
\draw [black, fill=blue!25] (A1) -- (A2) -- (C1) -- cycle;
\draw [black, fill=blue!25] (A4) -- (A1) -- (C1) -- cycle;
\draw [black, fill=blue!25, opacity=0.7] (A2) -- (A3) -- (B1) -- cycle;
\draw [black, fill=blue!25, opacity=0.7] (A3) -- (A4) -- (B1) -- cycle;
\draw [black, fill=blue!25, opacity=0.7] (A2) -- (A3) -- (C1) -- cycle;
\draw [black, fill=blue!25, opacity=0.7] (A3) -- (A4) -- (C1) -- cycle;
\node (A2) at (-1.75-0.9,0,0) {$(-1, 0, 0)$};
\node (A4) at (1.75+0.75,0,0) {$(1, 0, 0)$};
\node (B1) at (0,1.75+0.25,0) {$(0, 0, 1)$};
\node (C1) at (0,-1.75-0.25,0) {$(0, 0, -1)$};
\end{tikzpicture}
\end{minipage}
\end{tabular}
\caption{3-dimensional unit cube (left) and cross-polytope (right)}
\label{fig:example3}
\end{figure}
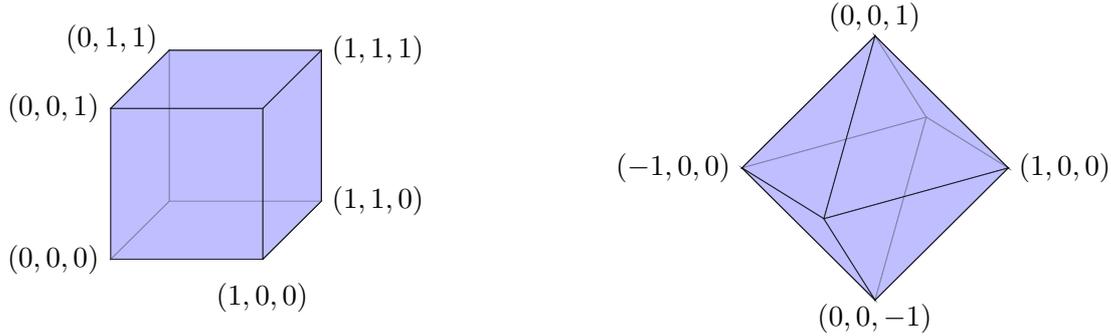

A more compact representation of the convex hull is  often useful from the viewpoint of computation.  For instance, the V-representation in Example \ref{cube} and the H-representation in Example \ref{cross} 
require more computer memory even for small $n$, whereas the H-representation in Example \ref{cube} and the V-representation in Example \ref{cross} need less memory even for  large $n$. Hence the H-representation in Example \ref{cube} and the V-representation in Example \ref{cross} are more suitable to deal with in actual computers when the dimension $n$ is large. 


\section{Application of the V-representation to model based calibration for automotive engines}\label{application}

We propose a way to  handle the V-representation of the convex hull of a set of points without conversion into the H-representation in Sections \ref{P1V} and \ref{P2V}. This way uses the results in \cite{pardalos}. Before mentioning them, we discuss the computational difficulty in using some MBC software in Section \ref{difficulty}.

\subsection{Computational difficulty due to the H-representation}\label{difficulty}

As we have already mentioned in Section \ref{bm}, the convex hull of a set of input  signals which make the automotive engine behave normally is one of the approximation of the AOD of the engine.  Let $V=\{v_1, \ldots, v_m\}$ be a set of input signals $v_1, \ldots, v_m\in\mathbb{R}^n$. Then the approximation via the convex hull is formulated as $\conv(V)$ and is the V-representation. On the other hand, for both \ref{P1} and \ref{P2} in Section \ref{bm},  it is converted into $P = \{v\in\mathbb{R}^n : Av \le b\}$ for some $A\in\mathbb{R}^{k\times n}$ and $b\in\mathbb{R}^{k}$ in some MBC software, such as \cite{mbctool}. This corresponds to the conversion of the V-representation  of the convex hull $\conv(V)$ into the H-representation, and after this conversion, \ref{P1} and \ref{P2} are {respectively} equivalent to
\begin{enumerate}[label=(P\arabic*)']
\item Problem of determining whether $A\hat{v}\le b$ or $A\hat{v}\not\le b$, and \label{P1'}
\item Solution of $\displaystyle\min_{v\in\mathbb{R}^n}\left\{f(v) : g_j(v) \ge 0 \ (j=1, \ldots, k), Av\le b\right\}$. \label{P2'}
\end{enumerate}


In general, the conversion is  computationally costly and generates too many half-spaces to be handled efficiently by actual   computers available RAM memory. This is the main computational difficulty in using the approximation of the AOD via the convex hull implemented in some MBC software. Table \ref{table:hrep} displays the computation time and the number of generated half-spaces for the conversion of the V-representation into the H-representation. In this numerical experiment\footnote{The specification on the used computer is as follows: OS is Ubuntu 14.04, the CPU is  Intel\textregistered \ Xeon\textregistered \  with  3.10GHz, and the memory is 128GB and version of  MATLAB is R2015b.}, we generated a set $V$ of $m$  points in $[-1, 1]^n$ randomly and used {\ttfamily vert2lcon.m} in \cite{vert2lcon}, which calls the built-in function {\ttfamily convexhulln} in MATLAB based on Qhull \cite{qhull}. ``--"  in Table \ref{table:hrep} {indicates} that we do not compute the conversion because it spends more than 1000 $\sec$. We observe from Table \ref{table:hrep}  that {when} $n$ is not so large, the conversion is not so computationally intensive and is rather fast. However, when $n$ is larger (typically more than 10), generating the convex hull in the H-representation becomes computationally intensive. Moreover, since it generates many half-spaces, we can expect that the optimization in \ref{P2'} will also be computationally intensive. 

\begin{table}
\centering
\begin{tabular}{|c|cccccc|}
\hline
&\multicolumn{6}{c|}{$n$}\\
$m$& 5 & 7 & 9 & 11 & 13&15\\
\hline
 50 & 0.34 & 0.15 & 1.23 & 11.25 & 72.05 & 391.27 \\
  &      566 &     5,084 &    42,430 &   279,804 &  1,517,292 &  6,898,066 \\
\hline
100 & 0.04 & 0.42 & 8.41 & 107.85 & 1506.60 & -- \\
 &     1,326 &    16,382 &   229,218 &  2,399,099 & 25,526,149 & -- \\
\hline
200 & 0.06 & 1.20 & 32.88 & 699.06 & -- & -- \\
 &     1,970 &    42,918 &   851,321 & 13,002,403 & -- & -- \\
\hline
1000 & 0.15 & 7.44 & 394.21 & -- & -- & -- \\
 &     6,724 &   238,486 &  8,053,847 & -- & -- & -- \\
\hline
2000 & 0.24 & 13.87 & 980.61 & -- & -- & -- \\
 &     9,262 &   427,048 & 17,550,631 & -- & -- & -- \\
\hline
\end{tabular}
\caption{Numerical results on the conversion of  the V-representation into the H-representation : computation time [$\sec$] (upper) and the number of generated half-spaces (lower)}
\label{table:hrep}
\end{table}%

\subsection{Application of the V-representation to \ref{P1} : to determine whether a new point is in the convex hull or not}\label{P1V}

Let $V=\{v_1, \ldots, v_m\}$ be a set of points in $\mathbb{R}^n$. For \ref{P1}  in Section \ref{bm}, {\itshape i.e.}, the problem of determining  whether $\hat{v} \in \conv(V)$ or $\hat{v} \not\in \conv(V)$,  we have two approaches via H-representation and V-representation. In the approach via H-representation, after converting $\conv(V)$ to the linear inequalities $Av\le b$, we need to check whether $A\hat{v}\le b$ or not.  It is relatively easy to check $A\hat{v}\le b$, while the conversion is computationally intensive for not so large $m$ and/or $n$ as in Table \ref{table:hrep}. On the other hand, in the approach via V-representation, the linear programming ({\itshape abbr.} LP) method is available. At the end of this subsection, we will show that  the approach via V-representation is much faster than the H-representation. 

Fundamentally, LP can be regarded as an optimization problem, {\itshape i.e.}, the problem of minimization or maximization of a linear objective function over a polyhedron. The simplex method and interior-point method are efficient algorithms to solve a LP problem or detect the infeasibility of the problem. In addition, {\ttfamily linprog} implemented in Optimization Toolbox offered by MathWorks and \cite{cplex}, are available as commercial software to solve LP problems. Refer to \cite{chvatal, luenbuger},  for more details on LP. 

One can determine whether a new point $\hat{v}$ is in $\conv(V)$ or not by solving the following LP problem:
\begin{equation}\label{convLP}
\displaystyle\min_{\alpha_1, \ldots, \alpha_m} \left\{\displaystyle\sum_{i=1}^m c_i \alpha_i : 
 \displaystyle\sum_{i=1}^m \alpha_i v_i = \hat{v},  \displaystyle\sum_{i=1}^m \alpha_i = 1, \alpha_i \ge 0 \ (i=1, \ldots, m)\right\}, 
\end{equation}
where $c\in\mathbb{R}^n$ is fixed {arbitrarily}. Since any convex combination of $\hat{v}$ with $v_1, \ldots, v_m$ is feasible in (\ref{convLP}), we see that 
\begin{itemize}
\item if the optimal value of (\ref{convLP}) is finite, then $\hat{v}$ is in $\conv(V)$, and, 
\item otherwise (\ref{convLP}) is infeasible, {\itshape i.e.}, the feasible region is empty, and thus $\hat{v}$ is not in $\conv(V)$. 
\end{itemize}
Hence one can determine whether a new point $\hat{v}$ is in $P$ or not by solving (\ref{convLP})  instead of constructing $Av\le b$ for the H-representation of $\conv(V)$. 

Table \ref{table:vrep} displays the computation time for the same sets $V$ of $m$ points in $\mathbb{R}^n$ as Table \ref{table:hrep}. Here we generate $\hat{v}\in[-1, 1]^n$ randomly. We used {\ttfamily linprog}  to solve all LP problems.   Comparing Table \ref{table:vrep} with Table \ref{table:hrep}, we see that the determination of $\hat{v}\in \conv(V)$ via LP method is much faster in computation time than the conversion into the H-representation of $\conv(V)$. This implies that  the H-representation for \ref{P1} will require more time to compute than the V-representation.  For instance, in the case $(m, n) = (1000, 9)$, the same $V$ is used in Tables \ref{table:hrep} and \ref{table:vrep}, and it spends 394.21  seconds to construct the H-representation of $\conv(V)$, whereas it spends only 0.09  seconds to determine whether $\hat{v}\in\conv(V)$ or not. Since we need to check $A\hat{v}\le b$ for the determination of $\hat{v}\in\conv(V)$ via the H-representation, where $A$ and $b$ are constructed by the H-representation of $\conv(V)$, the total amount of computation time via the H-representation is more than 394.21 seconds. Therefore, we can conclude from Tables \ref{table:hrep} and \ref{table:vrep} that the V-representation is less computationally intensive than H-representation.

\begin{table}[htbp]
\centering
\begin{tabular}{|c|cccccc|}
\hline
&\multicolumn{6}{c|}{$n$}\\
$m$& 5 & 7 & 9 & 11 & 13&15\\
\hline
 50 & 0.26 & 0.02 & 0.01 & 0.02 & 0.02 & 0.04 \\
\hline
100 & 0.03 & 0.01 & 0.01 & 0.01 & 0.02 & -- \\
\hline
200 & 0.02 & 0.01 & 0.02 & 0.13 & -- & -- \\
\hline
1000 & 0.02 & 0.03 & 0.09 & -- & -- & -- \\
\hline
2000 & 0.03 & 0.04 & 0.12 & -- & -- & -- \\
\hline
\end{tabular}
\caption{Computation time  [$\sec$]  to  determine whether a new point is in the convex hull or not by using LP method}
\label{table:vrep}
\end{table}%

\subsection{Application of the V-representation to \ref{P2} : an optimization problem in the frame of MBC response surface methodology}\label{P2V}

As we have already mentioned in \ref{P2} of Section \ref{bm}, the following optimization problems are typically solved by using MBC models obtained using  the response surface methodology:
\begin{equation}\label{rsm}
\min_{v\in\mathbb{R}^n}\left\{f(v) : g_j(v) \ge 0 \ (j=1, \ldots, k), v\in  P\right\}, 
\end{equation}
where $P$ is the approximation of AOD by the convex  hull for a set $V=\{v_1, \ldots, v_m\}$ of points in $\mathbb{R}^n$, {\itshape i.e.} $P=\conv(V)$. Since any $v\in \conv(V)$ can be represented by a convex combination of $v_1, \ldots, v_m$, the optimization (\ref{rsm}) can be equivalently reformulated as 
  \begin{equation}\label{rsm2}
\displaystyle\min_{\alpha_1, \ldots, \alpha_m\in\mathbb{R}}\left\{\tilde{f}(\alpha_1, \ldots, \alpha_m) : \begin{array}{l}
\tilde{g}_j(\alpha_1, \ldots, \alpha_m) \ge 0 \ (j=1, \ldots, k),\\
 \displaystyle\sum_{i=1}^m\alpha_i = 1, \alpha_i\ge 0 \ (i=1, \ldots, m)
 \end{array}
 \right\}, 
\end{equation}
where $\tilde{f}(\alpha_1, \ldots, \alpha_m) = f\left(\displaystyle\sum_{i=1}^m \alpha_i v_i\right)$ and $\tilde{g}_j$ is defined in a similar manner to $\tilde{f}$.

Before showing numerical comparison of (\ref{rsm2}) with (\ref{rsm}), we mention some advantages and disadvantages of the formulation (\ref{rsm2}): 
\begin{enumerate}[label=(\Roman*)]
\item One can skip the process of constructing the H-representation of $\conv(V)$. As we have already seen in Table \ref{table:hrep}, the conversion is computationally intensive, and thus one can greatly reduce the computational cost.  
\item Since one does not apply the conversion of the V-representation of $\conv(V)$ into the H-representation, the number of inequality constraints in (\ref{rsm2}) is much lower than (\ref{rsm}) formulated by the H-representation. Consequently, the feasibility check of a generated solution in algorithms of optimization for (\ref{rsm2}) is much easier than (\ref{rsm}). 
\item In contrast, the number of variables in (\ref{rsm2}) increases. In fact, it is $m$, while for (\ref{rsm}) is $n$, and thus the computational cost  increases in one evaluation of a function value at a given solution.  This is the disadvantage of the formulation (\ref{rsm2}). For instance, we will see in Table \ref{table:experiment} that  (\ref{rsm}) formulated by the H-representation is more efficient than (\ref{rsm2}) for $n = 4$.  
\end{enumerate}

To compare (\ref{rsm2}) with (\ref{rsm}) formulated by the H-representation, we use a diesel engine data set.  This data set consists of  875 observations and each measured observation   consists of following  nine  engine measurements, {\itshape i.e.}, Start of main injection event MAINSOI [$\deg$CA], Common rail fuel-pressure FUELPRESS [MPa], Variable-geometry turbo charger [VGT], vane position VGTPOS [mm], Exhaust gas recirculation (EGR) valve opening position EGRPOS [ratio], Amount of injected fuel mass during main injection event MAINFUEL[mg/stroke], Mass-flow ratio of recirculated exhaust gas EGRMF [ratio], Air-Fuel ratio AFR [ratio], VGT rotational speed VGTSPEED [rpm], and in-cylinder peak pressure PEAKPRESS [MPa]. The measurements were performed at seven specific engine operating points, expressed as (Engine Speed SPEED [rpm], Brake Torque BTQ [Nm]) pairs. 

Next, we generated point-by-point response surface models, {\itshape i.e.}, seven models, for the Brake-specific Fuel Consumption BSFC [g/kWh], by using the following three types of inputs from this diesel engine data set: 
\begin{enumerate}[label=(\Alph* type)]
\item BSFC$_p$(MAINSOI, FUELPRESS, 
VGTPOS, EGRPOS), \label{A}
\item BSFC$_p$(MAINSOI, FUELPRESS, 
VGTPOS, EGRPOS, MAINFUEL, EGRMF, AFR), {and} \label{B}
\item BSFC$_p$(MAINSOI, FUELPRESS, 
VGTPOS, EGRPOS, MAINFUEL, EGRMF, AFR, VGTSPEED, PEAKPRESS),  \label{C}
\end{enumerate}
where $p=1, \ldots, 7$.

The dimension $n$ of these data sets {is} $4, 7$ and $9$, respectively. We considered different $n$ in order to investigate the scalability of our proposed approach and to compare the computational cost with the H-representation of the convex hull. 


Next, for each data set, we have solved  the following seven optimization problems, one for each operating point set: 
\begin{equation*}\label{experiment}
\displaystyle\min_{v\in\mathbb{R}^n} \left\{
f_p(v) : v \in P
\right\} \ (p=1, \ldots, 7),  
\end{equation*}
where $P=\conv(V)$, and $V$ consists of a subset of the initial 875 $n$-dimensional vectors, since the approach we adopted is a point-by-point one. The measured points in each subset are unique. As an indication, each local model consisted of 125 of such measurements, and for each local model a corresponding convex hull was generated. Next, an optimization problems was considered. For this, we generate seven objective functions $f_p$ (for example using BSFC as the objective to be minimized) and do not use any extra constraint $g_j(v)\ge 0$ in this numerical experiment, except for the boundary model constraint itself. In conclusion, we have performed a point-by-point minimization problem for BSFC. 

Table \ref{table:experiment} displays numerical comparison of (\ref{rsm2}) with (\ref{rsm}) formulated by the H-representation of $P$. In this numerical experiment\footnote{The specification on the used computer is as follows: OS is Windows 7, the cpu is  Intel\textregistered \ Core\texttrademark \ i7 with  3.60GHz, and the memory is 32GB and version of  MATLAB is R2015b.}, we use  MBC Toolbox \cite{mbctool} and compare computation time of  (\ref{rsm2}) with (\ref{rsm}). The third and fourth columns in Table \ref{table:experiment} are the computation time of the conversion of $P$ into the H-representation and the total of computation time for seven types of optimization, respectively.  We do not describe the time in (\ref{rsm2}), but  ``--" in Table \ref{table:experiment} because we do not convert $P$ into the H-representation. We used {\ttfamily fmincon} with interior-point algorithm implemented in Optimization toolbox of MATLAB to solve both (\ref{rsm}) and (\ref{rsm2}). The optimization settings that were used to obtain the solution are listed in Table \ref{table:options}. For the settings not listed in Table \ref{table:options}  the defaults settings were used.

We observe the following from Table \ref{table:experiment}.
\begin{enumerate}[label=(\roman*)]
\item In \ref{A} and \ref{B}, {\itshape i.e.}, $n=4$ and $n=7$, (\ref{rsm}) formulated by the conversion of H-representation is faster than (\ref{rsm2}), whereas in \ref{C}, (\ref{rsm2}) is approximately 2 times faster than (\ref{rsm}). In fact, as we can expect form Table \ref{table:hrep}, the number of linear inequalities in (\ref{rsm}) considerably increases. Consequently, the evaluation of computed solutions at each iteration becomes computationally intensive.

\item The computation time of converting $\conv(V)$ into the H-representation considerably increases  as $n$ increases. This can be also expected from Table \ref{table:hrep}. As (\ref{rsm2}) can skip this conversion, we can expect that (\ref{rsm2}) is  more efficient than (\ref{rsm}) for $n\ge 9$.
\end{enumerate}

\begin{table}[htbp]
\centering
\begin{tabular}{|cc||c|c|}
\hline
&  & H-representation&Optimization\\
 \hline
(\ref{rsm})&\ref{A} & 0.06 &5.33\\
&\ref{B} & 7.72 &15.05\\
&\ref{C}  & 285.31 &66.65\\
\hline
(\ref{rsm2})&\ref{A}& -- &  45.52\\
&\ref{B} & -- & 26.06\\
&\ref{C} &-- & 37.94\\
\hline
\end{tabular}
\caption{Comparison of (\ref{rsm2}) with (\ref{rsm}) formulated by the H-representation of $P$ in computation time [$\sec$]}
\label{table:experiment}
\end{table}%

\begin{table}[htbp]
\centering
\begin{tabular}{|p{100mm}|c|}
\hline
Maximum number of function evaluations& 5000 for H-rep. \\
& 20000 for V-rep. \\
\hline
Maximum number of iterations& 500\\
\hline
Maximum change in variables for finite-difference gradients& 0.1\\
\hline
Minimum change in variables for finite-difference gradients&$10^{-8}$\\
\hline
Step tolerance for free variables &$10^{-6}$ \\
\hline
Constraint violation tolerance &$10^{-6}$ \\
\hline
Objective function tolerance&$10^{-6}$ \\
\hline
\end{tabular}
\caption{Optimization options used by {\ttfamily fmincon }}
\label{table:options}
\end{table}%
%
%

\section{Conclusion}\label{conclusion}
We propose a way to reduce the computational cost in the approximation of the AOD via the convex hull. The H-representation of the convex hull is identified as the main bottleneck. We focus on the two processes in MBC and observe  that the computational cost {is} greatly reduced when using the  V-representation of a set of points instead of  the H-representation. More precisely, when the dimension $n$ of the space in which a set $V$ of points lies is less than seven, the H-representation is not so computationally intensive. Otherwise it becomes more computationally intensive than the V-representation.

Enumeration of all the extreme points in $\conv(V)$ may be useful  when the V-representation is applied to \ref{P1} and \ref{P2} described in Section \ref{application}. In fact, this is ensured by Krein-Milman's theorem that for every bounded closed convex set $A$, $\conv(A) = \conv(\ext(A))$ holds. 
A simple way to enumerate all extreme points of $\conv(V)$ is to solve the following LP problem for every $v_k\in V$:
\begin{equation}\label{extLP}
\displaystyle\min_{\alpha_i \ (i\neq k)} \left\{\displaystyle\sum_{i\neq k}^m c_i\alpha_i : 
\displaystyle\sum_{i\neq k}^m \alpha_i v_i = v_k, \displaystyle\sum_{i\neq k}^m\alpha_i = 1, \alpha_i \ge 0 \ (i\neq k)\right\}. 
\end{equation}
If the optimal value is finite, then $v_k$ is not {an} extreme point in $\conv(V)$ because $v_k$ is a convex combination with other $v_i$ except for $v_k$. Otherwise (\ref{extLP}) is infeasible, and thus $v_k$ is an extreme point. This way is used as a pre-processing for \ref{P1} and \ref{P2}. If $\conv(V)$ consists a few extreme points in comparison to the set $V$, then we can expect the improvement of performance for \ref{P1} and \ref{P2}. See \cite{pardalos} for a much faster algorithm of the enumeration of all extreme points.

\section*{Acknowledgements}
This article is based on discussion in workshops of the IMI Joint Research Projects ``Research of Boundary Modeling" (Principal investigator : Satoru Watanabe) and ``Research of Engine Calibration from the Viewpoint of Mathematical Science" (Principal investigator : Masahiro Satou). We would like to thank the participants in these workshops for a fruitful discussion and secretaries for hospitality. We would also like to thank Mr. Akira Ohata in Technova Inc. and Prof. Hiroyuki Ochiai in Kyushu University for significant comments to improve the presentation of this article.

\end{document}